\documentclass[a4paper,fleqn,usenatbib]{mnras}

\usepackage{newtxtext,newtxmath}

\usepackage[T1]{fontenc}
\usepackage{ae,aecompl}
\usepackage{lineno,hyperref}
\usepackage{gensymb}
\usepackage{amsmath}
\usepackage{booktabs,caption}
\usepackage[flushleft]{threeparttable}
\usepackage[justification=centering]{caption}
\usepackage{xcolor}
\usepackage{graphicx}
\usepackage{multicol}
\usepackage{dcolumn}
\usepackage{multirow}
\usepackage{xspace}
\usepackage{mathrsfs}
\usepackage{xfrac}
\usepackage{braket}
\usepackage{xifthen}
\usepackage{rotating}
\usepackage{longtable}
\usepackage{chemformula}

\def\cratio{$^{12}$C/$^{13}$C}

\usepackage{graphicx}	
\usepackage{amsmath}	
\usepackage{amssymb}	

\title[DC$_3$N observations towards high-mass star-forming regions]{DC$_3$N observations towards high-mass star-forming regions}

\author[Rivilla et al.]{
V. M. Rivilla$^{1}$, 
L. Colzi$^{1}$,
F. Fontani$^{1}$,
M. Melosso$^{2}$,
P. Caselli$^{3}$,
L. Bizzocchi$^{3}$,
\newauthor
F. Tamassia$^{4}$,
and L. Dore$^{2}$
\\
$^{1}$ INAF/Osservatorio Astrofisico di Arcetri, Largo Enrico Fermi 5, I-50125, Florence, Italy\\
$^{2}$ Dipartimento di Chimica ``Giacomo Ciamician'', Universit\`a di Bologna, Via F.~Selmi 2, 40126 Bologna, Italy  \\
$^{3}$ Center for Astrochemical Studies, Max Planck Institut f\"ur extraterrestrische Physik Gie\ss enbachstra\ss e 1, 85748 Garching bei M\"unchen, Germany \\
$^{4}$ Dipartimento di Chimica Industriale ``Toso Montanari'', Universit\`a di Bologna, Viale del Risorgimento 4, 40136 Bologna, Italy \\ 
}

\begin{document}
\label{firstpage}
\pagerange{\pageref{firstpage}--\pageref{lastpage}}
\maketitle

\begin{abstract}

We present the study of deuteration of cyanoacetylene (HC$_3$N) towards a sample of 28 high-mass star-forming cores divided into different evolutionary stages, from starless to evolved protostellar cores. We report for the first time the detection of DC$_3$N towards 15 high-mass cores. The abundance ratios of DC$_3$N with respect HC$_3$N range in the interval 0.003$-$0.022, lower than those found in low-mas protostars and dark clouds. 
No significant trend with the evolutionary stage, or with the kinetic temperature of the region, has been found.
We compare the level of deuteration of HC$_3$N with those of other molecules towards the same sample, finding weak correlation with species formed only or predominantly in gas phase (N$_2$H$^+$ and HNC, respectively), and no correlation with species formed only or predominantly on dust grains (CH$_3$OH and NH$_3$, respectively).
We also present a single-dish map of DC$_3$N towards the protocluster IRAS 05358+3543, which shows that DC$_3$N traces an extended envelope ($\sim$0.37 pc) and peaks towards two cold condensations separated from the positions of the protostars and the dust continuum. The observations presented in this work suggest that deuteration of HC$_3$N is produced in the gas of the cold outer parts of massive star-forming clumps, giving us an estimate of the deuteration factor prior to the formation of denser gas.


\end{abstract}

 



\section{Introduction}

\label{sec:astro}

The deuteration level of interstellar molecules is a powerful tool to study the history of star-forming regions.
The formation of deuterated species is significantly enhanced during the pre-stellar phases of star-forming regions due to the combination of high gas densities ($>$10$^{4}$ cm$^{-3}$) and low temperatures (T$\leq$20 K), which favour the depletion of CO onto interstellar dust surfaces (\citealt{caselli1999,walmsley2004}). This enhances the gas-phase abundances of several ions (e.g. H$_3^+$), in absence of its main destroyer, CO. Then, the reaction of H$_3^+$ with HD, the main reservoir of deuterium in molecular clouds, produces H$_{2}{\rm D}^{+}$ through the reaction H$_{3}^{+}$ + HD $\rightarrow$ H$_{2}$D$^{+}$ + H$_{2}$ + 232 K,
when the species involved are in the para-H$_2$ form.
The exothermicity\footnote{The exotermicity of this reaction depends on the ortho-para ratio of the species involved, and that can vary between 84 and 256 K ( \citealt{millar1989,pagani1992}).} of this reaction prevents the backwards reaction at low temperatures, thus increasing the gas-phase abundance of H$_{2}{\rm D}^{+}$. This species transfers deuterium to other molecules efficiently, increasing the deuterium fractionation in molecules by several orders of magnitude with respect to the primordial value of D/H $\sim$ (1$-$2)$\times$10$^{-5}$ in the solar neighborhood (\citealt{oliveira2003,linsky2006}).
Therefore, the deuterated fraction, $D_{\rm frac}$, defined as the column density ratio of one species containing deuterium to its counterpart containing hydrogen, can be used as a useful probe of the evolution of star-forming cores.

In low-mass star-forming cores, \citet{caselli2002} found that $D_{\rm frac}$ significantly increases during the pre-stellar phase, when the core density profile becomes more centrally peaked due to the collapse. As a consequence, the CO freeze-out increases, along with the abundance of deuterated molecules in the gas-phase. Multiple observations have shown that the value of $D_{\rm frac}$ reaches its maximum in pre-stellar cores close to gravitational collapse (e.g. \citealt{crapsi2005}), and decreases during the later protostellar phases (\citealt{emprechtinger2009}).

In high-mass star-forming regions, how $D_{\rm frac}$ changes with the evolutionary stage is less clear. 
In previous works, we have studied the deuteration of different species towards a sample of high-mass star-forming cores at different evolutionary stages, from pre-stellar to protostellar phases. \citet{fontani2011} found that enhanced values of N$_2$D$^+$ are found towards high-mass pre-stellar cores, similarly to low-mass counterparts. However, other molecular species, such as HNC, NH$_3$ and CH$_3$OH, do not show any clear trend with the evolutionary stage using the same sample of sources (\citealt{fontani2014,fontani2015,colzi2018a}).

In this work, we focus on the carbon-chain molecule cyanoacetylene (HC$_3$N). Little is known about its deuteration, since only a few observations have been able to detect it in space. 
DC$_3$N was first detected by \citet{langer1980tmc} towards the low-mass dark cloud TMC-1, and by \citet{howe1994} towards several other dark clouds, finding values of $D_{\rm frac}$ in the range 0.03-0.13. In the protostellar phase, DC$_3$N has been detected only towards a handful of sources: IRAS 16293-2422 (\citealt{al2017iras}), L1527 (\citealt{sakai2009} and \citealt{araki2016}), Chamaeleon MMS1 (\citealt{cordiner2012}), and SVS13-A (\citealt{bianchi2019}), with values of $D_{\rm frac}$  $\sim$ 0.03$-$0.4.


There is still no unambiguous detection of DC$_3$N towards high-mass star-forming regions. Only two tentative detections have been reported towards Orion KL and the Sgr B2 N2 hot molecular cores (\citealt{esplugues2013orion,belloche2016emoca}, respectively). These observations, along with the non detection reported towards NGC 2264 CMM3 (\citealt{cordiner2012}), derived upper limits of D$_{\rm frac}<$ 0.015, which are lower than those reported in low-mass protostars. 

There are two main questions that need to be answered about deuterium fractionation of HC$_3$N.
First, is HC$_3$N deuterated through gas-phase reactions, like simple species such as N$_2$H$^+$ (e.g. \citealt{fontani2011}), or alternatively on the surface of intersellar dust grains via H-D exchange reactions, as suggested for other carbon-chain molecule (c-C$_3$H$_2$, \citealt{chantzos2018})?
%
And second, how $D_{\rm frac}$ changes with the evolutionary stage of the star-forming core? 
To properly answer these questions,
detections of DC$_3$N towards massive cores at different evolutionary stages are needed. With this aim, we present the first study of deuteration of HC$_3$N towards a sample of massive star-forming regions spanning a wide range of evolutionary stages, from initial starless cores to more evolved phases. We report the detection of the DC$_3$N $J$=11$-$10 transition towards 15 regions, and present the first maps of DC$_3$N in one of the sources of the sample, the high-mass protocluster IRAS 05358+3543.

\begin{table}
\tabcolsep 3.0pt
\centering
\caption{List of the observed sources. Cols. 2 and 3 give the equatorial coordinates of the sources. Cols. 4 and 5 give the source distance to the Sun, and the Galactocentric distance, respectively. For the references on distances from the Sun see \citet{fontani2011}. In Col. 6 the $^{12}$C/$^{13}$C ratios, derived using the galactocentric trend by \citet{milam2005}, and the associated uncertainties, are given.
In the last column the kinetic temperatures of the clumps derived by \citet{fontani2015} are listed. For the sources without a direct derivation of $T_{\rm kin}$ (indicated with $^{a}$), the average value for that evolutionary stage was taken.}
\begin{tabular}{c c c c c c c}
\hline
  Source & $\alpha({\rm J2000})$ &$\delta({\rm J2000})$  & $d$ & {\it D}$_{\rm GC}$ & \cratio & $T_{\rm kin}$ \\
& (h m s)& (\degr \arcmin \arcsec)  & (kpc) &(kpc) &  & (K) \\
\midrule 
  \multicolumn{7}{c}{HMSC (cold)}\\
  \midrule
 I00117-mm2 & 00:14:26.3  &+64:28:28.0        &   1.8   &    9.5   &  69$\pm$21               & $14$\\
G034-G2(mm2) & 18:56:50.0 & +01:23:08.0      &    2.9  &   6.3    &  50$\pm$17    & $16^{a}$ \\
 G034-F1(mm8) & 18:53:19.1 & +01:26:53.0    &    3.7  &  5.8    &  47$\pm$16      & $16^{a}$ \\
G034-F2(mm7) & 18:53:16.5 & +01:26:10.0    &    3.7  &   5.8    &  47$\pm$16       & $16^{a}$ \\
G028-C1(mm9) & 18:42:46.9 & -04:04:08.0      &    5.0  &   4.7    &  40$\pm$15    & $17$ \\
I20293-WC & 20:31:10.7 & +40:03:28.0     &    2.0  &  8.3    &  62$\pm$19      & $17$ \\
I22134-B & 22:15:05.8 & +58:48:59.0      &    2.6  &   9.5    &  69$\pm$21       & $17$ \\
 05358-D1 & 05:39:11.5 & 35:45:46.2   &  1.8 & 10.3 & 74$\pm$22  & 26 \\
 05358-D2  & 05:39:11.5  &  35:45:23.9  & 1.8 & 10.3 & 74$\pm$22   & 20  \\
\midrule 
  \multicolumn{7}{c}{HMSC (warm)}\\
  \midrule
   AFGL5142-EC & 05:30:48.7 & +33:47:53.0    &    1.8  &    10.3   &  74$\pm$22   & $25$\\
 05358-mm3 &05:39:12.5  &  +35:45:55.0     &    1.8 & 10.3  &  74$\pm$22      & $30$ \\  
 I22134-G &22:15:10.5  & +58:48:59.0     &    2.6  &   9.5    &  69$\pm$21       & $25$ \\
\midrule 
\multicolumn{7}{c}{HMPO}\\
\midrule
 I00117-mm1 & 00:14:26.1& +64:28:44.0      &   1.8   &  9.5    &  69$\pm$21                 &$20$\\
AFGL5142-mm & 05:30:48.0& +33:47:54.0     &    1.8  &   10.3   &  74$\pm$22   &$34$\\
05358-mm1  & 05:39:13.1 & +35:45:51.0     &    1.8  &   10.3   &  74$\pm$22   &$39$\\
18089-1732 & 18:11:51.4 & -17:31:28.0     &    3.6  &   5.0    &  43$\pm$15       &$38$\\
18517+0437 & 18:54:14.2 & +04:41:41.0     &    2.9  &   6.4    &  51$\pm$17   &$40^{a}$\\
 G75-core & 20:21:44.0 & +37:26:38.0     &    3.8  &    8.4    &  63$\pm$19     &$96$\\
I20293-mm1 & 20:31:12.8 & +40:03:23.0     &    2.0  &   8.3    &  62$\pm$19     &$36$\\
I21307  & 21:32:30.6 & +51:02:16.0   &    3.2  &     9.3  &  68$\pm$20         &$21$\\
 I23385  &23:40:54.5  & +61:10:28.0   &    4.9  &    11.4   &  81$\pm$23      &$37$\\
  \midrule
\multicolumn{7}{c}{UC HII}\\
\midrule
G5.89-0.39 & 18:00:30.5& -24:04:01.0       &    1.3  &    7.2   &  55$\pm$18     & $32^{a}$\\
I19035-VLA1 &19:06:01.5& +06:46:35.0        &    2.2  &    7.0   &  54$\pm$18    &$39$\\
19410+2336  &19:43:11.4 & +23:44:06.0        &    2.1  &   7.7    &  58$\pm$18 & $21$\\
ON1 &20:10:09.1 & +31:31:36.0  &    2.5  &    8.0   &  60$\pm$19                  & $26$\\
I22134-VLA1 & 22:15:09.2 & +58:49:08.0    &    2.6 & 9.5  &  69$\pm$21        & $47$\\     
23033+5951  & 23:05:24.6&  +60:08:09.0    &    3.5  & 10.2  &  74$\pm$21       & $25$\\     
NGC7538-IRS9 & 23:14:01.8& +61:27:20.0     &    2.8  &    9.9 & 72$\pm$21  & $32^{a}$\\
\bottomrule
\end{tabular}
\label{tab-sample}
\end{table}

\begin{figure*}
\includegraphics[scale=0.31]{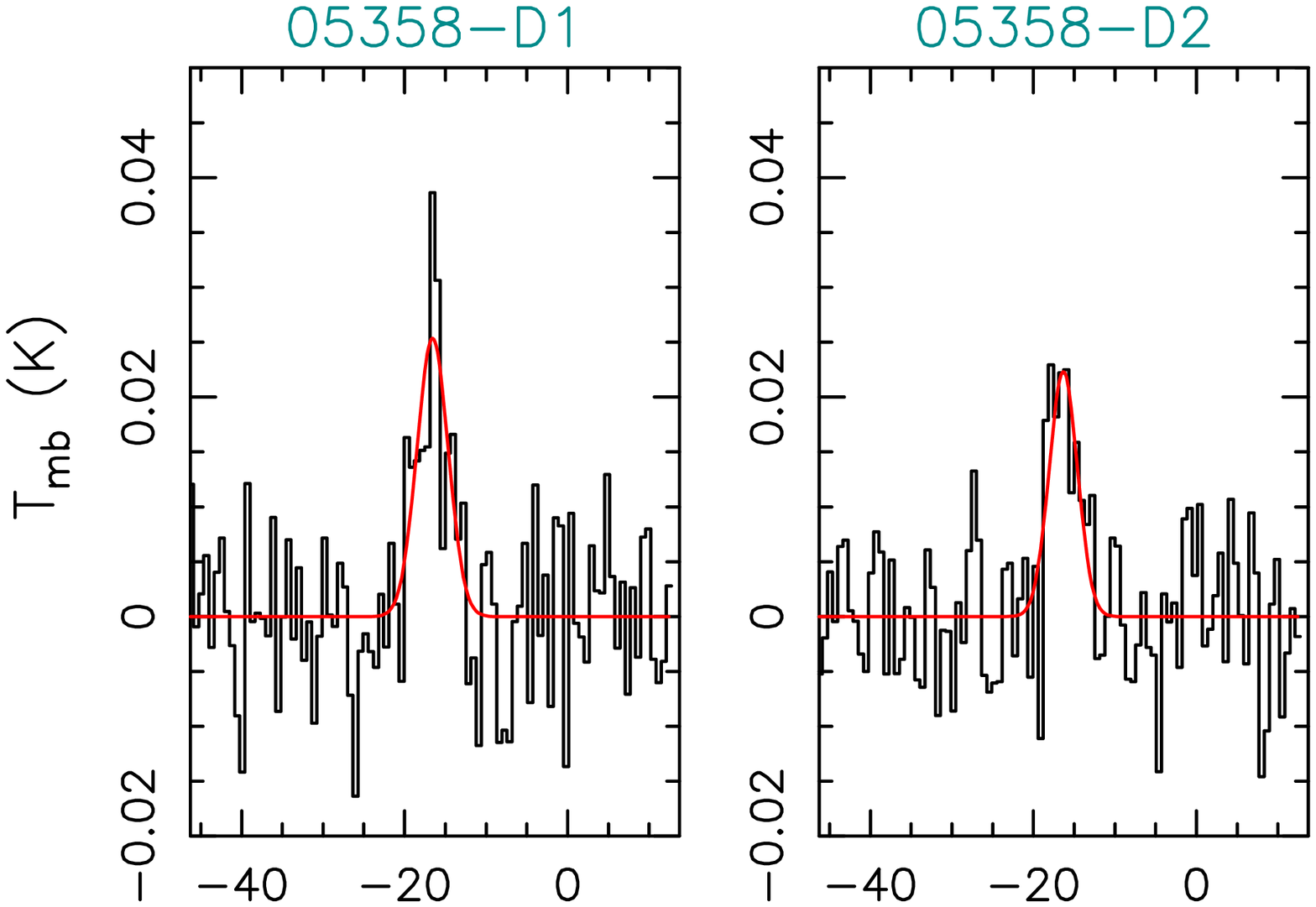}
\hspace{4mm}
\vspace{3mm}
\includegraphics[scale=0.31]{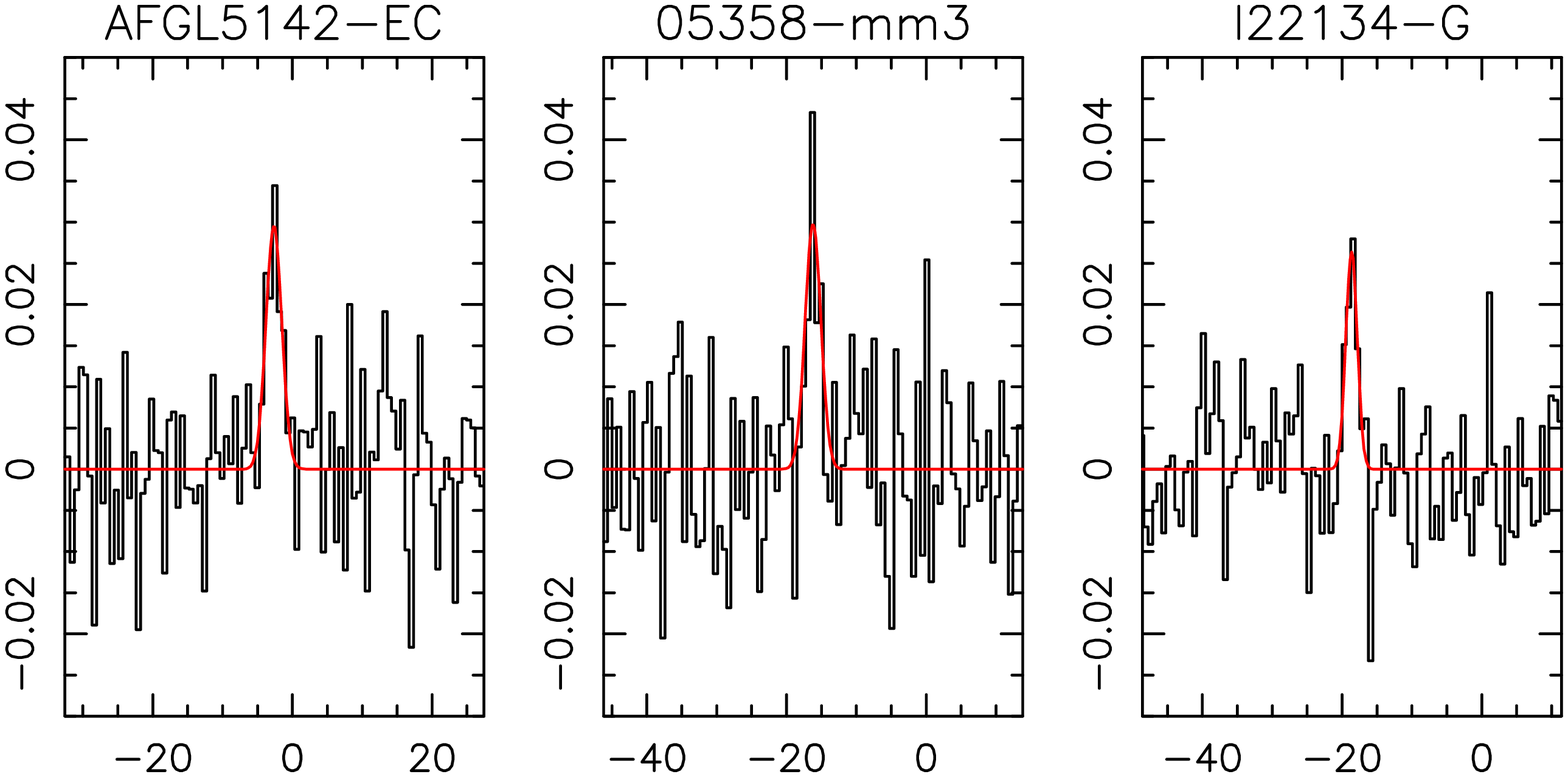}
\vspace{3mm}
\includegraphics[scale=0.31]{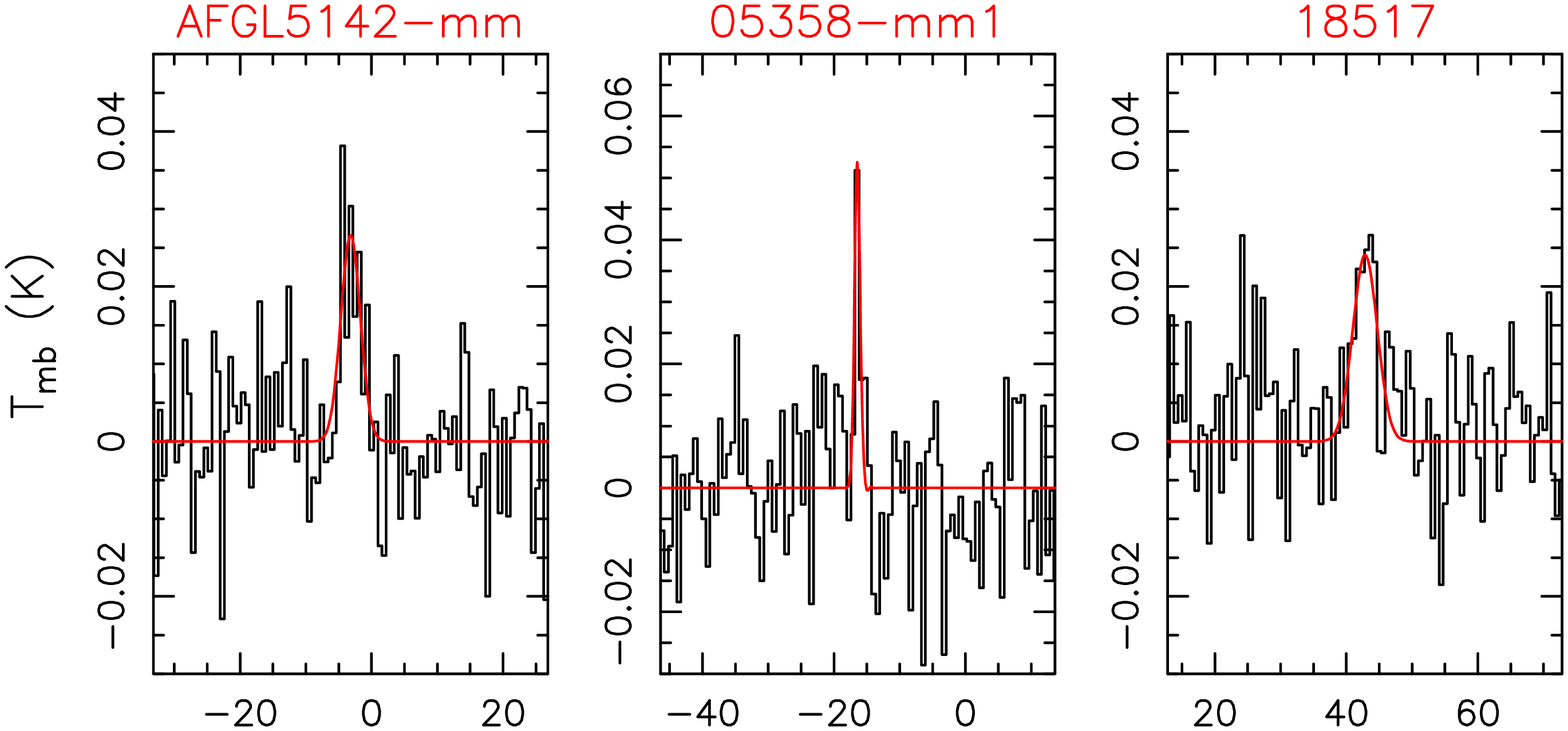}
\includegraphics[scale=0.31]{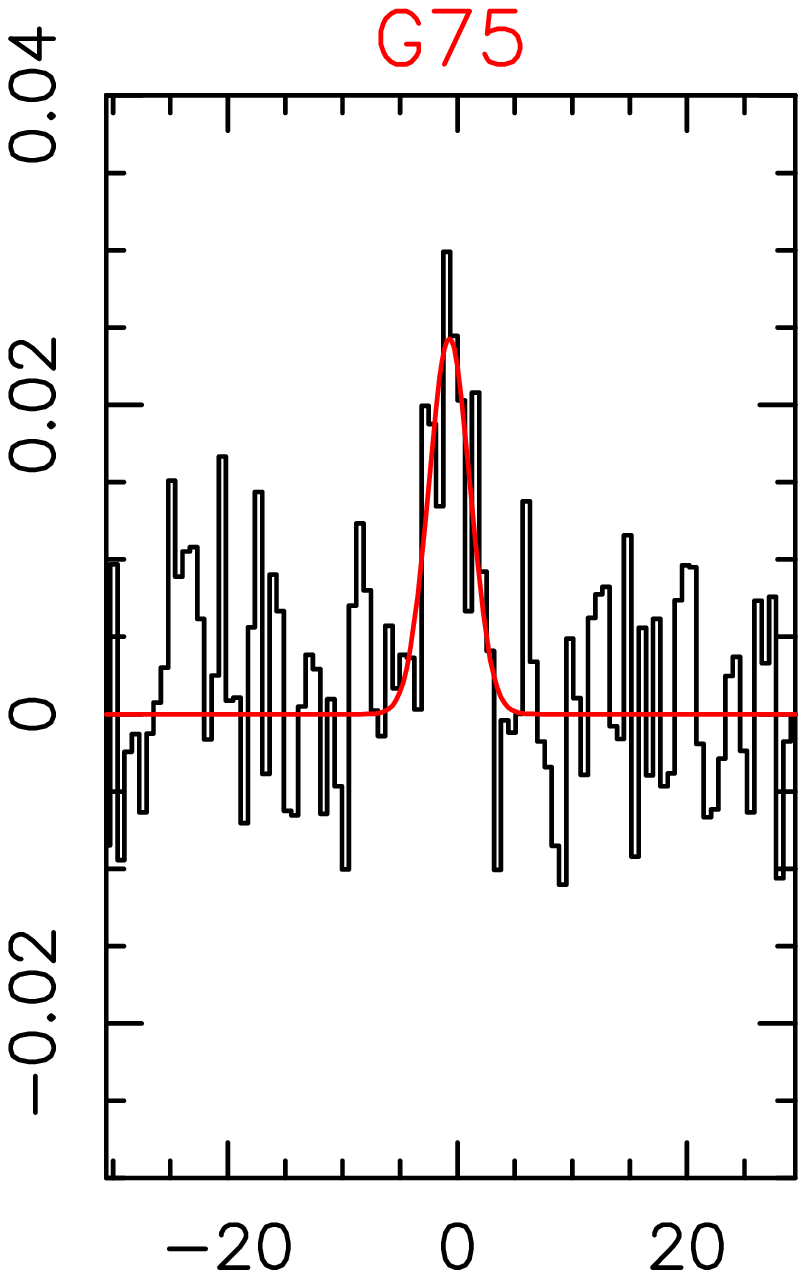}
\includegraphics[scale=0.31]{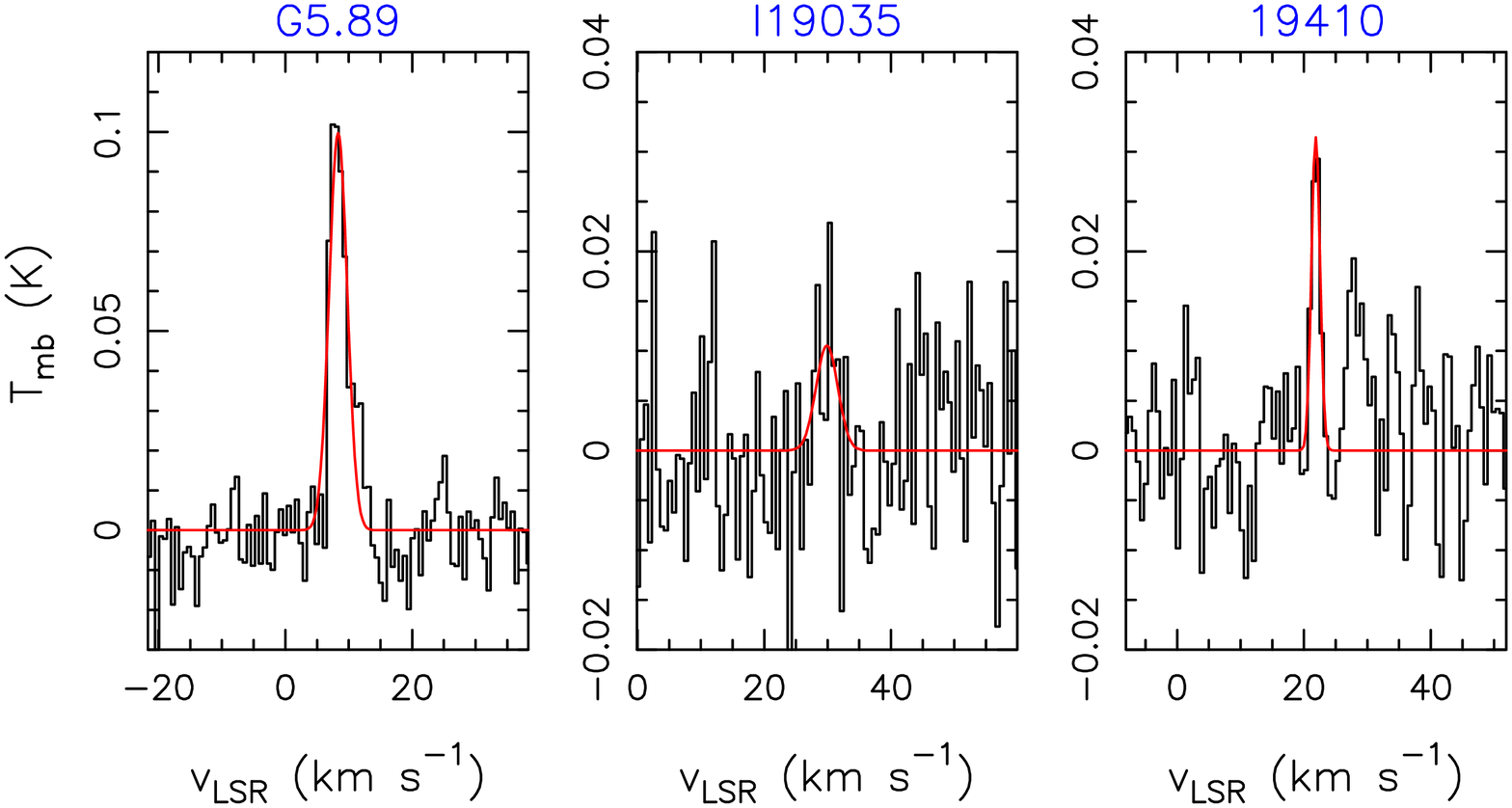}
\hspace{2mm}
\includegraphics[scale=0.31]{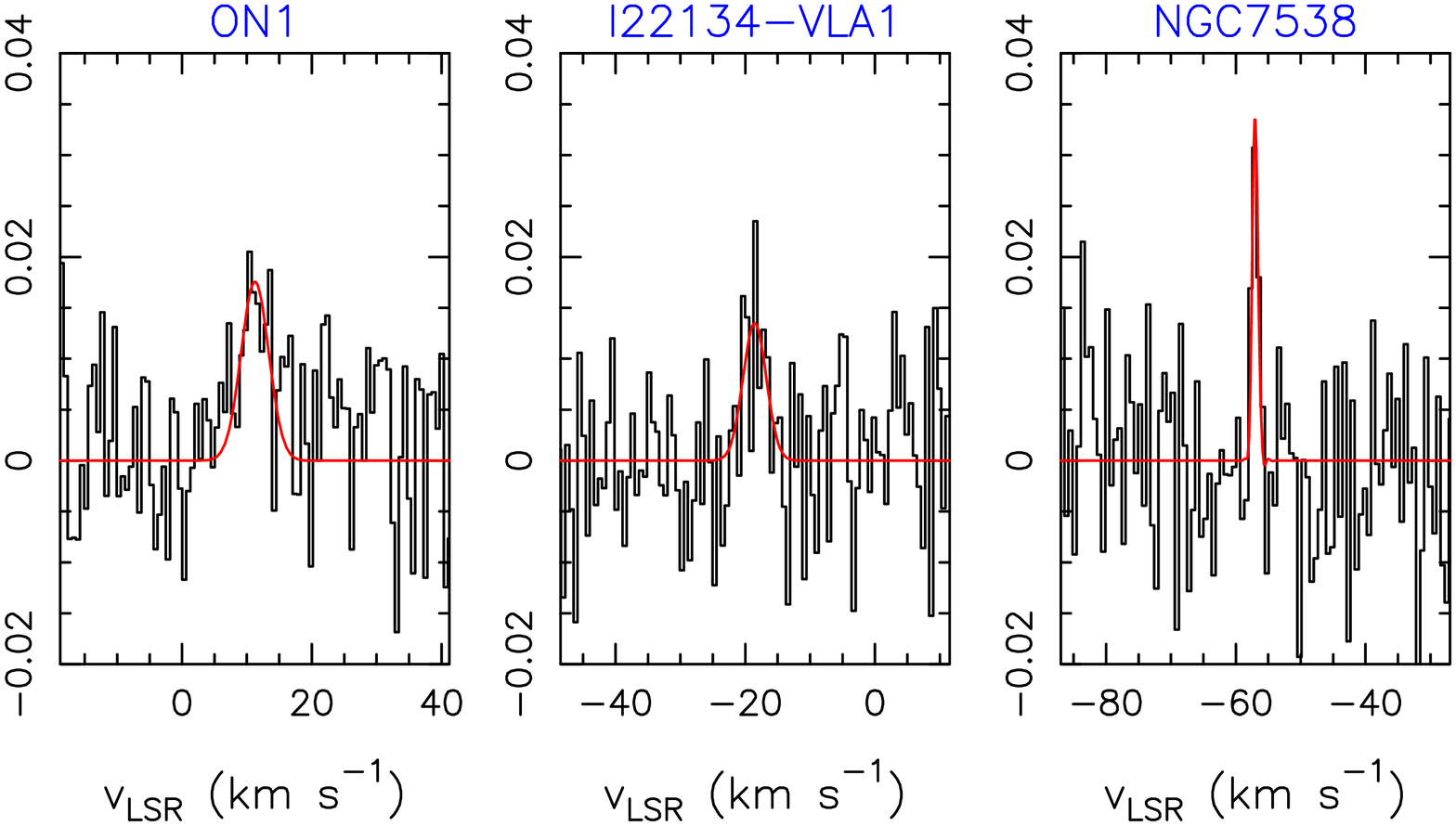}
\centering
\caption{DC$_3$N $J$=11$-$10 detections towards the sample of high-mass star-forming regions studied in this work. The observed IRAM 30m telescope spectra are shown with a black histogram, while the best LTE fits are shown with red curves. Each panel is centered at the systemic velocity of each source. The name of each source is indicated above each panel, coloured by evolutionary groups: dark green for {\it{cold}} HMSCs, black for {\it{warm}} HMSCs, red for HMPOs and blue for UC HII regions.}  
\label{fig-spectra}
\end{figure*}

\section{Observations}
\label{sec-observations}
We conducted astronomical observations using the 30m-diameter telescope of the Institut de Radioastronomie Millim\'etrique (IRAM), located at Pico Veleta (Granada, Spain), as part of the projects 129$-$12 and 040-19. In the project 129$-$12, we performed single-pointing observations towards 26 high-mass star-forming cores (see details about the sample in Sect. \ref{sec-sample}). We used the broad-band Eight MIxer Receiver (EMIR), covering $\sim$8 GHz from 89.11 to 96.89 GHz, and the fast Fourier Transform Spectrometer (FTS, \citealt{klein2012}) in a mode that provides a channel width of 195 kHz ($\sim$0.63 km s$^{-1}$). 
The observed setup includes the $J$=11$-$10 transition of DC$_3$N at 92.872373 GHz (spectroscopy from the recent laboratory measurements of Melosso et al. {\it in prep.}), and the $J$=10$-$9 transitions of two of the three $^{13}$C-isotopologues of HC$_3$N, HC$^{13}$CCN and HCC$^{13}$CN, at 90.593059 and 90.601777 GHz, respectively (\citealt{creswell1977,thorwirth2001}).
The spatial resolution of the observations, given by the half-power beam width of the antenna (HPBW), can be calculated as HPBW(\arcsec)=2460/$\nu$(GHz), which gives 26.5\arcsec at the frequecny of the DC$_3$N $J$=11$-$10 transition.
The spectra were calibrated using the GILDAS\footnote{The GILDAS software is is available at http://www.iram.fr/
IRAMFR/GILDAS} software package. The antenna temperature units, $T_{\rm A}^{*}$, were converted to main beam brightness temperature, $T_{\rm mb}$, via the relation $T_{\rm A}^{*}$ = $T_{\rm mb}\times$(B$_{\rm eff}$/F$_{\rm eff}$), using the corresponding telescope efficiencies\footnote{http://www.iram.es/IRAMES/mainWiki/Iram30mEfficiencies}. The flux density calibration uncertainty of the observations is $\sim$20$\%$. More details of the observations (e.g. weather conditions, background subtraction, focus and pointing) are presented in \citet{fontani2015}.

In the 040-19 project, we mapped with multiple telescope pointings the molecular emission of one of the star-forming regions of the sample, the high-mass protocluster IRAS 05358+3543 (hereafter 05358).
We also observed the $J$=11$-$10 transition of DC$_{3}$N, and the $J$=10$-$9 transitions of HC$^{13}$CCN and HCC$^{13}$CN, during observations performed in the periods 31st July to 5th August 2019 and 18th to 19th September 2019. As for the single-pointing observations, we used the 3mm receiver EMIR, and the FTS spectrometer with a frequency resolution of 195 kHz ($\sim$0.63 km s$^{-1}$). 
The observations were made in position-switching mode. Pointing was
checked every 1.5 h, and focus was corrected at the beginning of the observations, at dawn and every 4-6 h. 
The molecular datacubes were produced from On-The-Fly (OTF) mapping, covering an area of 120\arcsec $\times$ 120\arcsec (corresponding to $\sim$1 pc $\times$ 1 pc,
at the source distance of 1.8 kpc; \citealt{heyer1996}), with a central position of RA(J2000) = 05h:39m:13s.0, DEC(J2000)= 35°:45\arcmin:51\arcsec.
The integration time for each OTF map (vertical + horizontal + calibration times) was of $\sim$20 minutes.
The antenna temperature $T_{\rm A}^{*}$ was converted to main beam temperature $T_{\rm mb}$ by using the same expression given above. Llux calibration uncertainties of 20$\%$ are considered in the analysis. The GILDAS packages CLASS and GREG were used to reduce and post-process the data. Baselines were all fitted by constant functions or polynomials of order 1. We built the molecular data cube convolving the OTF map with a Gaussian kernel, using a regularly spaced grid with pixel size of 9\arcsec.



\section{Source sample}
\label{sec-sample}

The source sample, already used in several works (\citealt{fontani2011,fontani2014,fontani2015,colzi2018a,colzi2018b,mininni2018}), includes high-mass star-forming cores spanning a wide range of evolutionary stages: 

\begin{itemize}
\item High-Mass Starless Cores (HMSCs), which are not directly associated with indicators of on-going star formation, such as embedded infrared sources, outflows, or masers. As in previous works, we have divided this group into two subgroups: ${\it cold}$ and ${\it warm}$. The latter includes three regions (AFGL5141-EC, 05358-mm3 and I22134-G) that have temperatures $T_{\rm kin}\geq$25 K, and that can be externally heated by nearby protostellar objects detected by high-angular resolution observations (\citealt{zhang2002,busquet2011,sanchez-monge2011,colzi2019}). We note that our observations of these three {\it warm} HMSCs are likely contaminated by the emission of nearby protostellar sources, which fall in the IRAM 30m beam of 26.5\arcsec. This will be discussed in Section \ref{sec-discussion}.
We have added to the {\it cold} HMSC group two new cores that have been identified in the DC$_3$N maps of the 05358 region presented in this work (Sect. \ref{sec-maps}): 05358-D1 and D2.
\item High-Mass Protostellar Objects (HMPOs), which are associated with molecular outflows, and/or faint radio continuum emission (S$_{\rm \nu}$ at 3.6 cm $<$ 1 mJy) likely tracing a radio-jet, and/or infrared sources.
\item Ultracompact (UC) HII regions; associated with a strong radio-continuum emission (S$_{\rm \nu}$ at 3.6 cm $\geq$ 1 mJy), which likely traces photoionised gas by nascent massive stars.
\end{itemize}

All the selected sources are located at distances $d\leq$ 5 kpc.  The full list of the observed sources is shown in Table \ref{tab-sample}. More details about how the sources were selected and classified in the different evolutionary stages are given in \citet{fontani2011}. In summary,
we observed 28 high-mass star-forming cores divided into: 9 {\it{cold}} HMSCs, 3 {\it{warm}} HMSCs, 9 HMPOs, and 7 UC HII regions.



\begin{figure}
\includegraphics[scale=0.42]{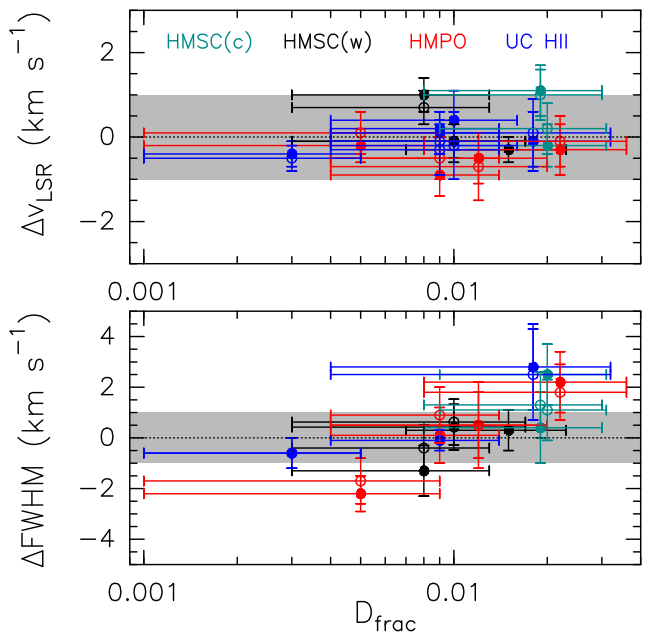}
\caption{Differences between the velocities (upper panel) and FWHM (lower panel) of DC$_3$N and the $^{13}$C-isotopologues. Filled(empty) circles correspond to HCC$^{13}$CN(HC$^{13}$CCN). The different colors indicate the different evolutionary groups, as indicated. The shaded area denotes the region $\pm$1 km s$^{-1}$.}
\label{fig-vel-FWHM}
\end{figure}

\begin{figure*}
\includegraphics[scale=0.35]{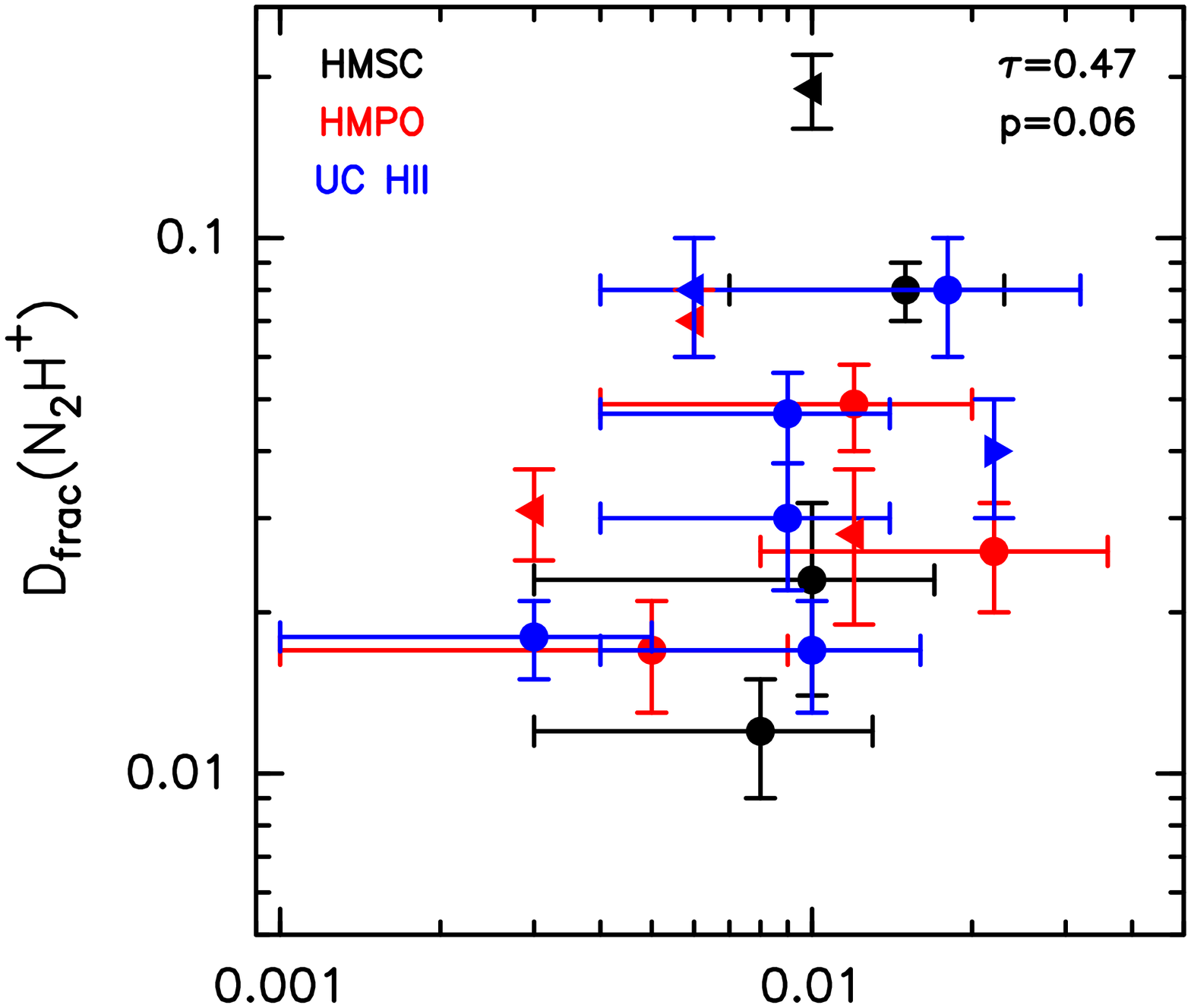}
\hspace{5mm}
\vspace{5mm}
\includegraphics[scale=0.35]{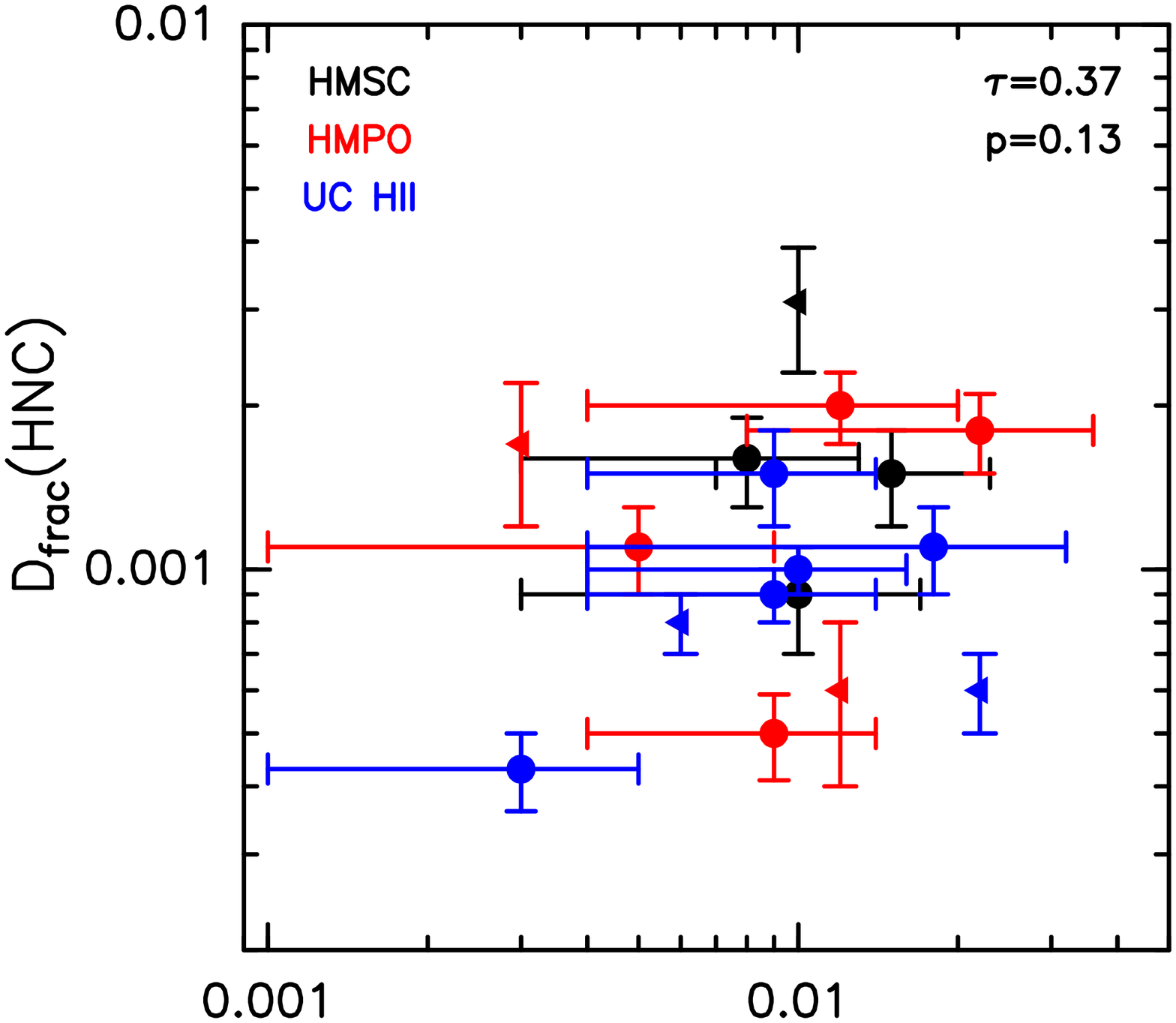}
\includegraphics[scale=0.35]{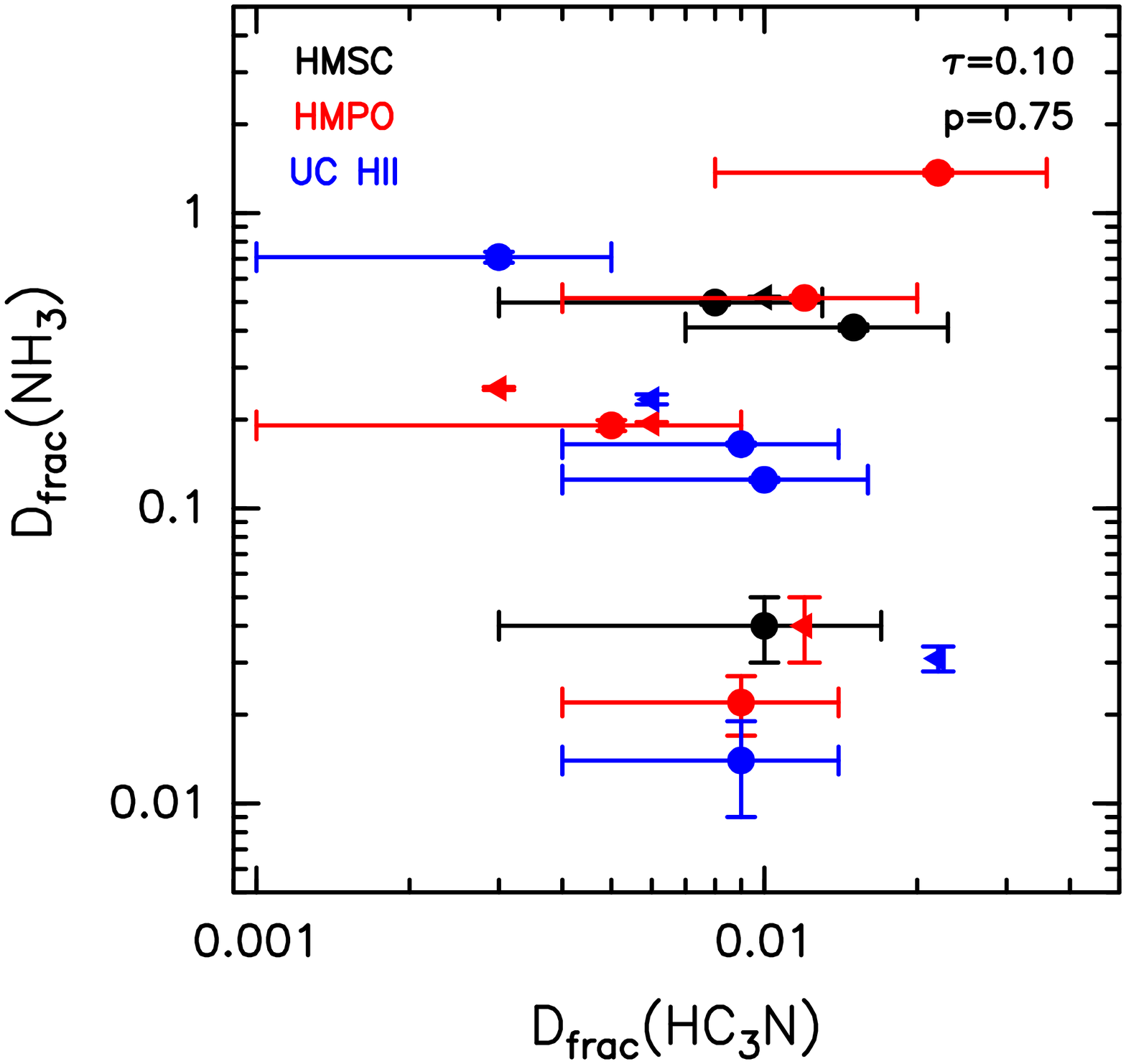}
\hspace{5mm}
\includegraphics[scale=0.35]{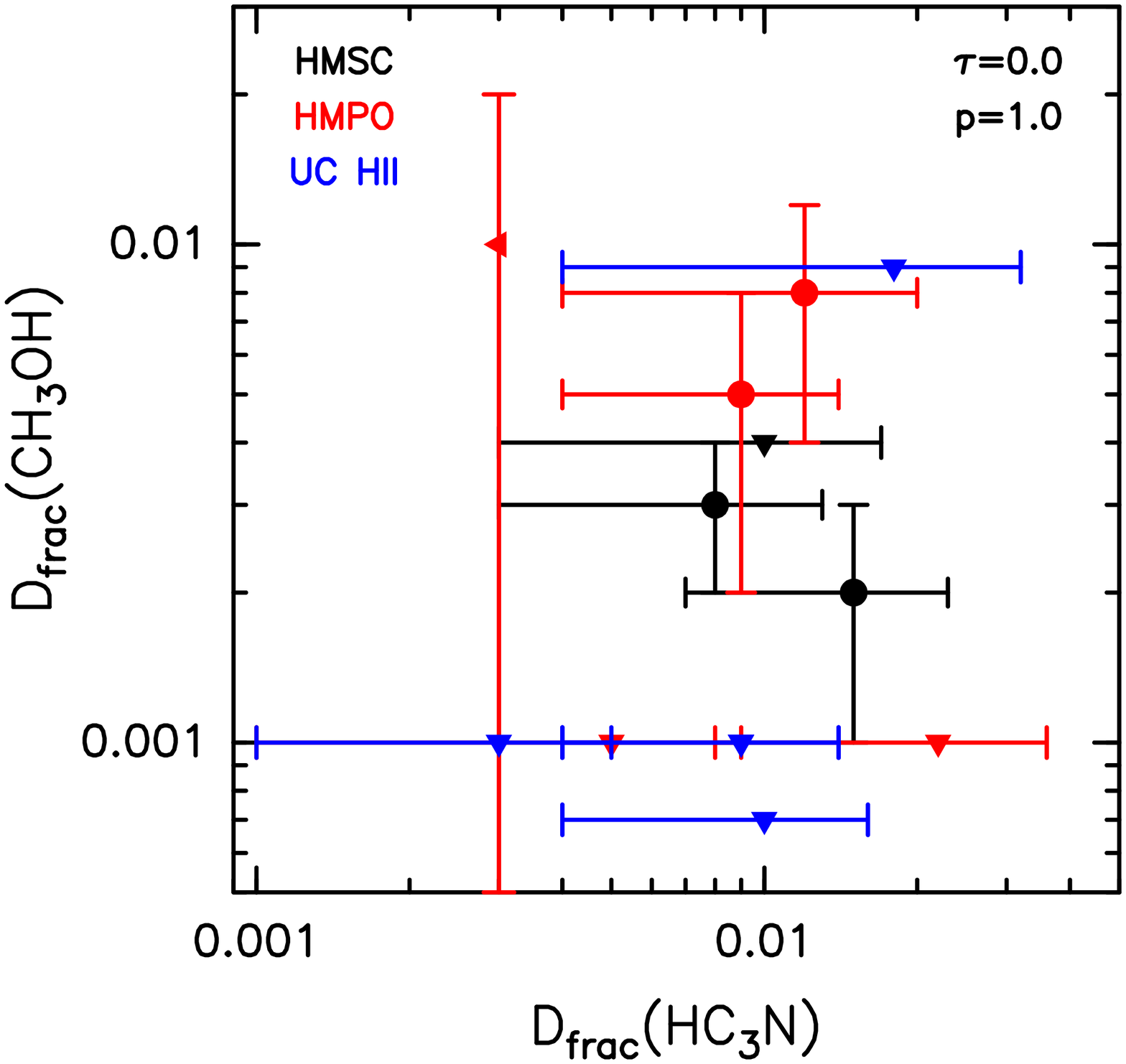}
\centering
\caption{Values of $D_{\rm frac}$ derived for HC$_3$N in the sample of high-mass star-forming regions studied in this work ({\it warm} HMSC, HMPO and UC HII regions), compared with those derived for other species in previous works. The different colours of the circles correspond to different evolutionary groups, as indicated in each panel. Upper limits are denoted by triangles. The outputs of the Kendall $\tau$ test ($\tau$ and double-side $p$ value) are indicated in the upper right corner of each panel.}
\label{fig-correlations}
\end{figure*}

\section{Data analysis and results}
\label{sec-results}

\subsection{Single-pointing observations}

The spectra were imported into the MADCUBA package{\footnote{Madrid Data Cube Analysis on ImageJ is a software developed at the Center of Astrobiology (CAB) in Madrid; http://cab.inta-csic.es/madcuba/Portada.html.}} (\citealt{martin2019}).
The identification of the molecular lines was performed using the SLIM (Spectral Line Identification and Modeling) tool within MADCUBA.
SLIM generates synthetic spectra of molecular species under the assumption of Local Thermodynamic Equilibrium (LTE) conditions.
We have implemented into SLIM the new DC$_3$N spectroscopic data recently obtained in the laboratory by Melosso et al. ({\it in prep.}). 

We detected the DC$_3$N $J$=11-10 transition towards 15 regions.  The transitions observed, shown in Figure \ref{fig-spectra}, include 2 {\it{cold}} HMSCs, 3 {\it{warm}} HMSCs, 4 HMPOs and 6 UC HII regions.
To obtain the best fit of the observed transitions we used the MADCUBA-AUTOFIT tool that compares the observed spectra with LTE synthetic spectra, and provides the best non-linear least-squares fit using the Levenberg$-$Marquardt (LM) algorithm (see details in \citealt{martin2019}). The free parameters of the fit are: column density of the molecule ($N$), excitation temperature ($T_{\rm ex}$), velocity (v$_{\rm LSR}$) and full width at half-maximum ($FWHM$) of the Gaussian profiles of the lines. We assumed that the molecular emission fills the beam of the telescope, and thus we did not apply any beam dilution correction. This assumption is reasonable, since the low energies of the transitions studied (E$_{\rm up}$=24-27 K), which suggest that these molecules trace relatively extended gas. In the particular case of the 05358 region, the emission maps we present in  Sect. \ref{sec-maps} confirm that this assumption is fulfilled.
Since we only have one rotational transition, we fixed the value of $T_{\rm ex}$ to the kinetic temperature $T_{\rm kin}$ of the cores (Table \ref{tab-sample}). We note that the column density of DC$_3$N can vary by factors $<$ 1.3(1.7) depending of the $T_{\rm ex}$ assumed in the range 15$-$50 K (15$-$100 K). 
The other parameters ($N$, v$_{\rm LSR}$ and $FWHM$) were left free whenever possible. Only if the LM algorithm did not converge we fixed v$_{\rm LSR}$ and/or $FWHM$ to values that visually match the observed lines. The resulting fits are shown in red overplotted to the observed spectra in Figure \ref{fig-spectra}, and the resulting parameters are presented in Table \ref{table-dfrac}.
In the case of non detections, upper limits for the column density of DC$_3$N were derived with SLIM, which uses the formula 3$\sigma\times\Delta$v/$\sqrt{N_{\rm chan}}$, where $\sigma$ is the {\it rms} noise of the spectra, and $N_{\rm chan}$ is the number of channels covered by the linewidth $\Delta$v. The derived upper limits for the column density of DC$_3$N
are shown in Table \ref{table-dfrac}.

To obtain the value of $D_{\rm frac}$, we derived the HC$_3$N column densities towards each source with the same procedure used for DC$_3$N. 
Since HC$_3$N is an abundant molecule in high-mass star-forming regions, its emission is expected to be optically thick, which prevents a reliable accurate derivation of its column density. For this reason, we opted to analyze the two $^{13}$C-isotopologues of HC$_3$N available in the observed frequency band (HC$^{13}$CCN and HCC$^{13}$CN), which are expected to be optically thin. 
We have used the spectroscopic entries of the Cologne Database for Molecular Spectroscopy (CDMS, \citealt{muller2001,muller2005,endres2016}), based on the laboratory works by \citet{creswell1977} and \citet{thorwirth2001}.
We calculated the average column density value of the $^{13}$C-isotopologues, and assumed the $^{12}$C/$^{13}$C ratio as a function of the Galactocentric distance found by \citet{milam2005}  (Table \ref{tab-sample}) to convert to HC$_3$N column density. The derived values of $D_{\rm frac}$ are presented in the last column of Table \ref{table-dfrac}. We note that, unlike the values of the column densities, the values of $D_{\rm frac}$ are nearly independent of the assumed $T_{\rm ex}$, whenever DC$_3$N and the $^{13}$C-isotopologues of HC$_3$N have a similar $T_{\rm ex}$.
We found values of $D_{\rm frac}$ in the range 0.003-0.022. For the sources with no DC$_3$N detection but HC$^{13}$CCN and HCC$^{13}$CN detections, we give the upper limits of $D_{\rm frac}$. Only in one case, the UC HII region I19035-VLA1, DC$_3$N was detected but not the $^{13}$C-isotopologues. For this source we show in Table \ref{table-dfrac} the derived lower limit of $D_{\rm frac}$, $\geq$0.022. 

To make sure that the $^{13}$C-isotopologues arise from similar gas than DC$_3$N, we have checked that their velocities and linewidths are similar. We show in Figure \ref{fig-vel-FWHM} that the differences of the velocities ($\Delta$v$_{\rm LSR}$) and linewidths ($\Delta FWHM$) between DC$_3$N and the $^{13}$C-isotopologues are always within a narrow range of $\pm$1 km s$^{-1}$, considering the uncertainties derived by the fits. This similarity in the kinematics supports our assumption that DC$_3$N and the $^{13}$C-isotopologues are likely tracing similar gas.

To compare the deuteration of HC$_3$N with that of other molecules already studied in previous works towards the same sample, we show in Figure \ref{fig-correlations} the $D_{\rm frac}$(HC$_3$N) versus those derived for N$_2$H$^{+}$ (\citealt{fontani2011}), HNC (\citealt{colzi2018a}), NH$_3$  (\citealt{fontani2015}) and CH$_3$OH (\citealt{fontani2015}).
We performed Kendall's $\tau$ tests (\citealt{kendall1938}) to search for possible correlations between the set of values of $D_{\rm frac}$ of the different species.
We considered only the sources in which the deuterated species have been detected (no upper limits).
The initial guess (null hypothesis) is that the two datasets are not correlated. The  Kendall's $\tau$ correlation parameter can adopt values between -1 and 1: is equal to 1 if a perfect correlation is found, 0 if there is no correlation (initial guess fulfilled), and -1 if there is perfect anticorrelation. As an output of the test we also give the 2-sided $p$ value, which gives the level of significance of the test between 0 and 1: 0 when it is statistically significant and 1 when it is totally not significant. 
The results of the tests are shown in the upper right corner of each panel in Figure \ref{fig-correlations}. The deuteration of HC$_3$N shows better correlations with  N$_2$H$^{+}$ and HNC rather than NH$_3$ and CH$_3$OH. There are (weak) correlations with N$_2$H$^{+}$ and HNC ($\tau$=0.47 and 0.37, respectively), which are statistically significant (6$\%$ and 13$\%$ probability that the initial guess is correct, i.e., no correlation). For NH$_3$ and CH$_3$OH, there is no correlation ($\tau$=0.1 and 0.0, respectively), with 75$\%$ and 100$\%$ probability of no correlation, respectively. We discuss the implications of these results in Sect. \ref{sec-discussion}.

\begin{figure*}
\includegraphics[scale=0.385]{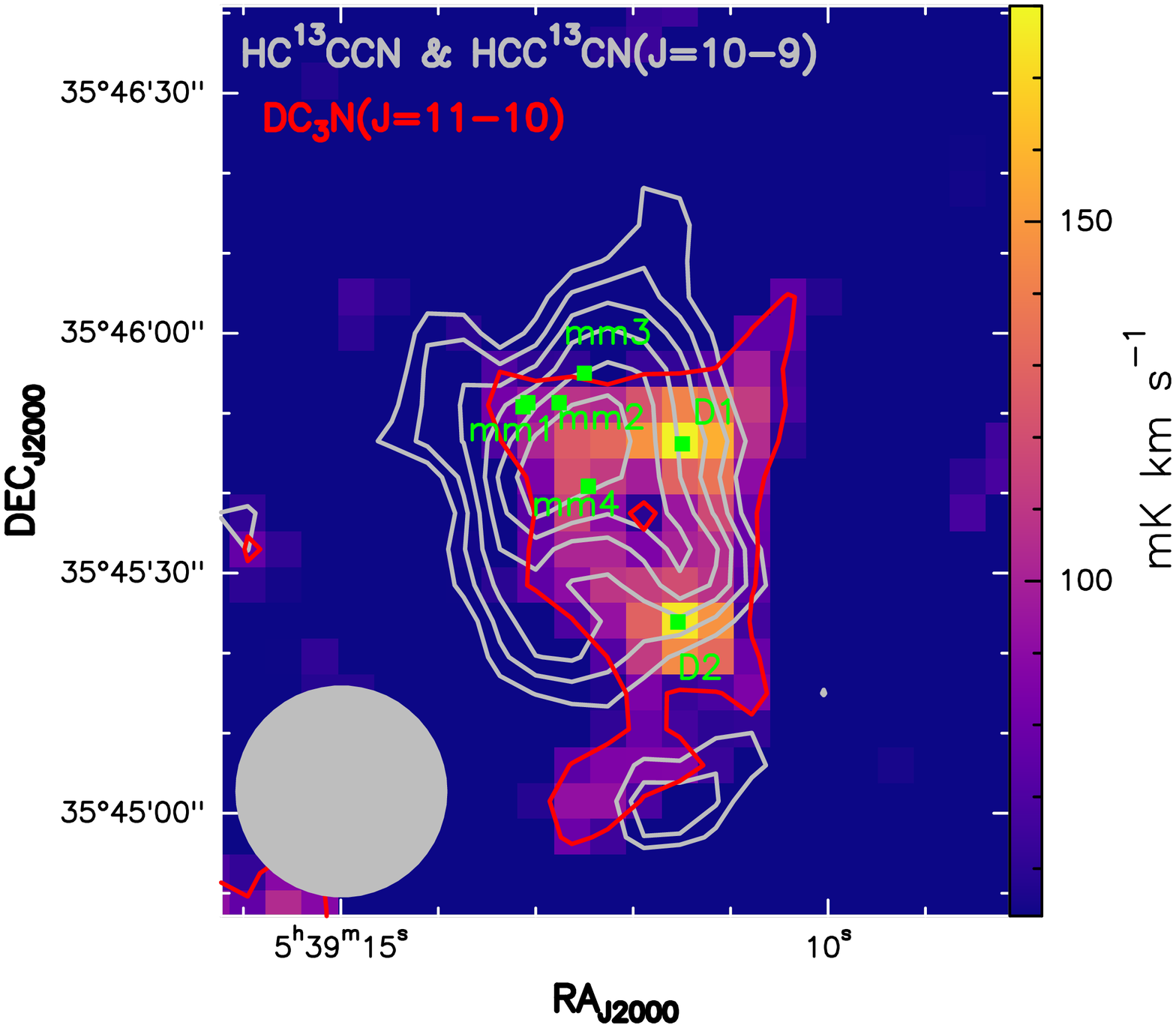}
\hspace{7mm}
\includegraphics[scale=0.385]{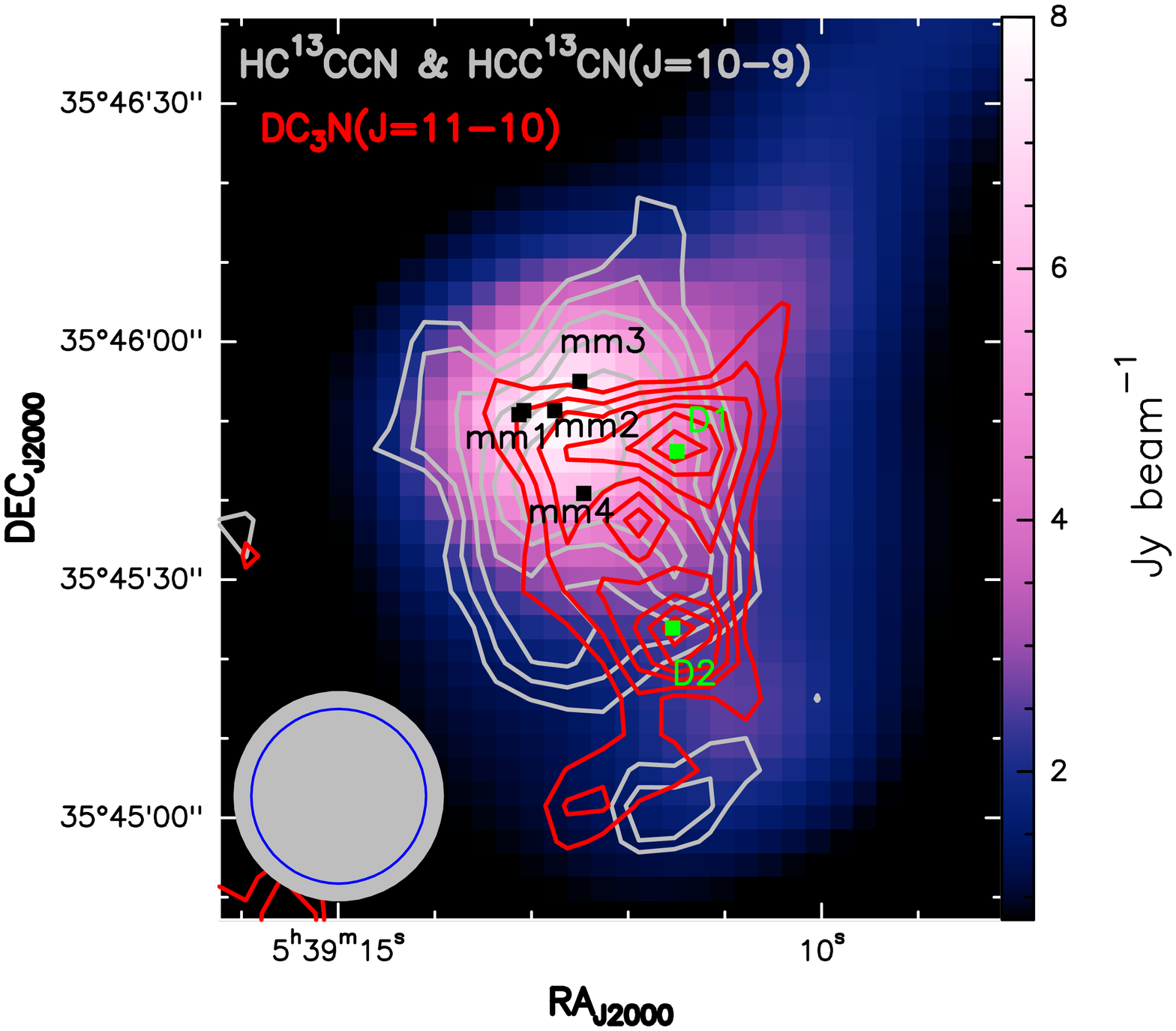}
\centering
\caption{{\it Left:} The color scale shows the integrated emission map of the $J$=11$-$10 transition of DC$_{3}$N.
The red contour correspond to 0.4 times the emission peak (178 mK km s$^{-1}$).
The gray contours show to the combined integrated emission of the $J$=10$-$9 transitions of HC$^{13}$CCN and HCC$^{13}$CN (0.9, 0.8, 0.7, 0.6, 0.5 and 0.4 times the emission peak, which is 265 mK km s$^{-1}$). 
The green squares indicate the positions of the continuum sources mm1, mm2, mm3 and mm4 (\citealt{colzi2019}), and the DC$_{3}$N peak positions (D1 and D2), respectively. The gray filled circle in the lower left corner indicate the beam of the DC$_3$N observation.
{\it Right}: The color scale shows the SCUBA 850 $\mu$m continuum observations, from \citet{difrancesco2008}. The beam size of the SCUBA map (22.9$\arcsec$) is indicated in the lower left corner with a blue open circle overplotted to the IRAM 30m beam (gray filled circle). The gray contour levels are the same as in the left panel, while the red contour levels are 0.9, 0.8, 0.7, 0.6, 0.5 and 0.4 times the peak of the integrated emission map of DC$_3$N.}
\label{fig-DC3Nmap}
\end{figure*}

\subsection{Maps of the 05358+3543 protocluster}
\label{sec-maps}

We produced the integrated maps of the $J$=11$-$10 transition of DC$_3$N, and the combined integrated map of the $J$=10$-$9 transitions of the two $^{13}$C-isotopologues (HC$^{13}$CCN and HCC$^{13}$CN), which are shown in Figure~\ref{fig-DC3Nmap}. 
We also indicate in the figure the positions of the continuum sources previously identified in the region with interferometric observations (\citealt{colzi2019}): mm1, mm2, mm3 and mm4. 
For comparison, we also show in the right panel of Figure \ref{fig-DC3Nmap} the continuum map at 850 $\mu$m obtained with the Submillimetre Common User Bolometer Array (SCUBA) of the 15m-diameter James Clerk Maxwell Telescope (JCMT), from \citet{difrancesco2008}.
The SCUBA observations have a HPBW of 22.9$\arcsec$, similar to our IRAM 30m observations. 

The main protostellar activity in the 05358 protocluster is located in the mm1/mm2/mm3 region, which matches with the SCUBA continuum peak,
and where multiple molecular outflows have been observed (\citealt{beuther2002}).
The source mm1 corresponds to one of the HMPOs studied in the previous section, and exhibits compact molecular emission with hot core chemistry (\citealt{leurini2007}; and Colzi, {\it priv. communication}).
The source mm3 is one of the {\it warm} HMSCs of our sample, and it is considered to be pre-stellar because no compact line emission has been detected in this position. The evolutionary state of mm2 and the recently discovered mm4 (\citealt{colzi2019}) are still unknown. As shown in the Figure \ref{fig-DC3Nmap}, the positions of mm1 and mm3 are separated by only 8.3\arcsec, while the beam of the IRAM 30m observations is $\sim$26\arcsec. Therefore, as previously discussed, it is likely that the spectra from the {\it warm} HMSC in mm3 is contaminated by the nearby protostellar activity. 

Figure~\ref{fig-DC3Nmap} shows that the integrated emission of DC$_3$N and the $^{13}$C-isotopologues of HC$_3$N are extended, larger than the IRAM 30m beam. To measure the area of the emitting regions we have considered the contour in which the emission falls to a  40$\%$ of the peak emission. The emission area of DC$_3$N and the $^{13}$C-isotopologues are $\sim$1446\arcsec$^2$ and $\sim$2042\arcsec$^2$, respectively. These emitting areas correspond to those of circular regions with diameters 0.37 pc and 0.44 pc, respectively.
In both cases the sizes are larger than that of the IRAM 30m beam ($\sim$0.26 pc). 
Therefore, the assumption of emitting region filling the telescope beam used in the previous section is fulfilled, at least, in the case of the 05358 region.



The integrated map of the $^{13}$C-isotopologues of HC$_3$N peak in the region located between mm1/mm2 and mm4, slightly shifted with respect to the peak of the SCUBA continuum by $\leq$10\arcsec. This shift is likely not significant, since it is less than half of the beams of the molecular and continuum observations, and considering some uncertainty of the pointings of the different telescopes.

Interestingly, the morphology of the emission of DC$_3$N is not fully coincident with that of the $^{13}$C-isotopologues. Figure~\ref{fig-DC3Nmap} shows that the overall DC$_3$N emission is shifted towards SW, with two main peaks, hereafter D1 and D2, not coincident with the peak of the $^{13}$C-isotopologues nor with the SCUBA continuum peak. Since no signs of active star formation towards D1 and D2 have been reported, we considered them as {\it{cold}} HMSCs, and added them to the sample of sources (Table \ref{tab-sample}). From the datacube, we extracted the spectra of circular regions centered at the D1 and D2 positions with diameters matching the IRAM 30m beam. We performed the analysis of DC$_3$N,  HC$^{13}$CCN and HCC$^{13}$CN using the same procedure described in the previous section. The values of $T_{\rm kin}$ were derived from NH$_3$ (as for the other cores of the sample), using the VLA interferometric maps from \citet{lu2014}, and integrating in a circular area with a diameter matching the IRAM 30 beam. The values ontained for D1 and D2 are 26 and 20 K, respectively, lower than that considered for mm1 and mm3 (39 K and 30 K, respectively; Table \ref{tab-sample}), which supports their {\it{cold}} starless nature. The derived values of $D_{\rm frac}$ are 0.020$\pm$0.011 and 0.019$\pm$0.011 for D1 and D2, respectively (Table \ref{table-dfrac}). These values are higher that those found towards mm1 and mm3, 0.005$\pm$0.02 and 0.008$\pm$0.03, respectively.




\begin{table}
\tabcolsep 3.25pt
\caption{Total DC$_3$N column density and D-fractionation of DC$_{3}$N. Values without errors are fixed in the fit procedure.}
\label{table-dfrac}
\begin{tabular}{c c c c c }
 \hline
 Source & $N_{\rm tot}$ & v$_{\rm LSR}$  & $\Delta$v & $D_{\rm frac}$(HC$_{3}$N)\\
& ($\times$10$^{11}$ cm$^{-2}$) & (km s$^{-1}$) & (km s$^{-1}$) &  \\
 \midrule
  \multicolumn{5}{c}{HMSC (cold)}\\
  \midrule
 I00117-mm2           &    $\leq$1.6        &      --            &  --            &  --  \\
G034-G2(mm2)          &      $\leq$1.4      &      --            &  --            &  --  \\
 G034-F1(mm8)         &       $\leq$1.3     &      --            &  --            &  --  \\
G034-F2(mm7)          &      $\leq$1.3      &      --            &  --            &  --  \\
G028-C1(mm9)          &    $\leq$1.7        &      --            &  --            &  --  \\
I20293-WC             &      $\leq$1.4      &      --            &  --            &  $\leq$0.010      \\
I22134-B              &     $\leq$1.3       &      --            &  --            &  --  \\
 05358-D1 &       5.4$\pm$1.4   & -16.6$\pm$0.4 & 4.5$\pm$0.9  &  0.020$\pm$0.011 \\
 05358-D2  &     4.6$\pm$1.2  & -16.3$\pm$0.4 &  4.1$\pm$0.8 &  0.019$\pm$0.011  \\
 \midrule
  \multicolumn{5}{c}{HMSC (warm)}\\
 \midrule
  AFGL5142-EC          &     3.7$\pm$0.9    &      -2.6$\pm$0.2  &  2.7$\pm$0.5   &  0.015$\pm$0.004  \\
 05358-mm3            &     3.1$\pm$0.9    &      -16.2$\pm$0.2 &  1.9$\pm$0.5   &  0.008$\pm$0.003 \\ 
  I22134-G    &  2.4$\pm$0.8     &      -18.6$\pm$0.3 &  1.9$\pm$0.6   &  0.010$\pm$0.005  \\
  \midrule
\multicolumn{5}{c}{HMPO}\\
\midrule
 I00117-mm1           &     $\leq$1.7       &     --             &  --           &  --  \\
AFGL5142-mm           &     4.5$\pm$1.3    &     -3.2$\pm$0.4   &  3.5$\pm$0.9  &  0.012$\pm$0.005  \\
05358-mm1             &     2.3$\pm$0.8    &     -16.4$\pm$0.2  &  0.9$\pm$0.3  &  0.005$\pm$0.002 \\
18089-1732            &     $\leq$1.9       &     --             &  --           &  $\leq$0.003  \\
18517+0437            &      5.2$\pm$1.4   &     42.9$\pm$0.4  &   4.4$\pm$0.8 &  0.022$\pm$0.006 \\
 G75-core             &       9$\pm$3      &     -0.7$\pm$0.3   &   4.3$\pm$0.7 &  0.009$\pm$0.002  \\
I20293-mm1            &      $\leq$1.4      &     --             &  --           &  $\leq$0.003 \\
I21307                &    $\leq$1.6        &     --             &  --           &  --  \\
 I23385               &     $\leq$1.1       &     --             &  --           &  $\leq$0.012  \\
  \midrule
\multicolumn{5}{c}{UC HII}\\
\midrule
G5.89-0.39            &     16$\pm$4       &    8.3$\pm$0.2    &  3.4$\pm$0.4  &  0.0030$\pm$0.0005 \\
I19035-VLA1           &      2.1$\pm$0.9   &    30.0$\pm$0.5     &  4$\pm$1      & $\geq$0.022  \\
19410+2336            &     2.4$\pm$0.6    &    21.9$\pm$0.1   &  1.6$\pm$0.2  &  0.009$\pm$0.002 \\
ON1        &      4$\pm$1       &    11.2$\pm$0.5   &  5            &   0.010$\pm$0.003\\
I22134-VLA1           &       2.9$\pm$1.1  &    -18.5$\pm$0.6   &  3.9$\pm$1.4  &   0.018$\pm$0.009\\     
23033+5951            &      $\leq$1.1      &    --              &  --           &  $\leq$0.006 \\     
NGC 7538-IRS9         &      1.6$\pm$0.5   &    -57.1$\pm$0.2   &  1            &   0.009$\pm$0.002\\
\hline
\end{tabular}
 \end{table}

\begin{figure}
\includegraphics[scale=0.375]{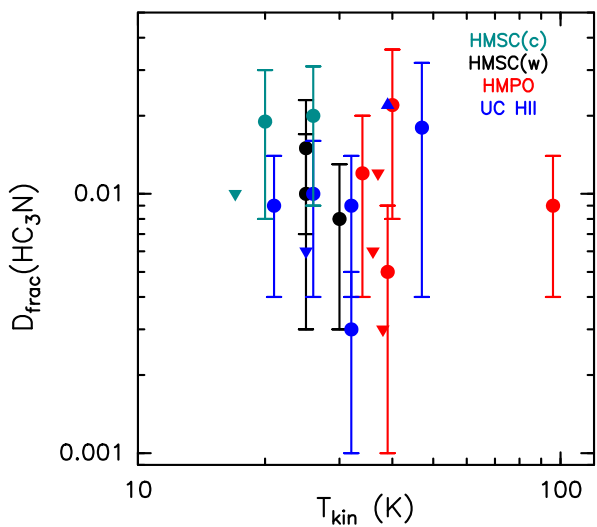}
\centering
\caption{$D_{\rm frac}$ of HC$_3$N as a function of the kinetic temperature $T_{\rm kin}$ towards the sample of high-mass star-forming regions studied in this work. The different colours of the circles correspond to different evolutionary groups, as indicated in the upper right corner. Upper limits of $D_{\rm frac}$ are denoted by triangles pointing downwards, while the only lower limit is denoted by the triangle pointing upwards.}
\label{fig-DC3N-Tkin}
\end{figure}

\section{Discussion}
\label{sec-discussion}

It is still not clear how DC$_3$N is formed in star-forming regions.
\citet{langer1980tmc} suggested several ion-molecule reactions that can form it during the cold pre-stellar phase:

\begin{subequations}
\label{eq-gas1}
\begin{align} 
{\rm    H_2D^+ + HC_3N \rightarrow HDC_3N^+ + H_2~,} \\
{\rm    HDCN^+ + C_2H_2 \rightarrow HDC_3N^+  + H_2~,} \\
{\rm    H_2CN^+  + C_2HD \rightarrow  HDC_3N^+  + H_2~,}
\end{align}
\end{subequations}
and

\begin{equation} 
\label{eq-gas2}
{\rm HDC_3N^+ + e \rightarrow DC_3N + H~. }  
\end{equation}
%

Other possible chemical pathway might be directly linked to the formation of HC$_3$N itself. Many observational works, from the study of its $^{13}$C isotopologues, have suggested that HC$_3$N is formed through the gas-phase neutral-neutral reaction between acetylene (C$_2$H$_2$) and CN (e.g. \citealt{takano1998,taniguchi2016,li2016,araki2016,taniguchi2017}). Therefore, this pathway might also contribute to form DC$_3$N via: 

\begin{equation} 
\label{eq-c2h2}
{\rm C_2HD + CN \rightarrow DC_3N + H~. } 
\end{equation}

An alternative route would imply surface chemistry through D-H exchange reactions after the freeze-out of HC$_3$N, as proposed by \citet{chantzos2018} for other carbon-chain species like cyclopropenylidene (c$-$C$_3$H$_2$). 
A fourth possibility was suggested by \citet{esplugues2013orion}, based on gas-phase hot chemistry during the protostellar phase. 

We discuss in the following how our detections of DC$_3$N towards a sample of massive cores can help us to understand how and when HC$_3$N is deuterated during high-mass star formation.
The comparison of $D_{\rm frac}$ of different species in our sample (Figure \ref{fig-correlations})
shows that HC$_3$N is better correlated with N$_2$H$^+$ and DNC than with NH$_3$ and CH$_3$OH. Interestingly, these species are formed only (the former) and predominantly (the latter) in gas phase. The opposite case is represented by CH$_3$OH and NH$_3$, which are produced efficiently on grain surfaces through hydrogenation of N and CO, respectively. The (weak) correlation of HC$_3$N with the species formed in gas phase, and the null correlation with species formed on grains, supports that the deuteration of HC$_3$N is likely produced through gas-phase chemistry rather than surface chemistry.


We have not found a clear correlation between the presence of DC$_3$N and the evolutionary stage, since we detected it in 2 {\it{cold}} HMSCs, 3 {\it{warm}} HMSCs, 4 HMPOs and 6 UC HII regions. Supporting this lack of dependence with the evolutionary phase, we do not find a correlation between $D_{\rm frac}$ and the kinetic temperature $T_{\rm kin}$ of the sources (Figure \ref{fig-DC3N-Tkin}).
We note that although the non detections of DC$_3$N in the sample correspond mainly to {\it{cold}} HMSCs (Table \ref{table-dfrac}), this does not imply necessarily that deuteration is lower in this early evolutionary stage. These non detections are likely due to sensitivity limits of the observations, since in most of the cases, the $^{13}$C-isotopologues are also not detected. We have checked that the 6 {\it{cold}} HMSCs with no detection of any of the isotopologues of HC$_3$N are those with weaker line intensities in other molecular species. As an example, the N$_2$H$^+$ observations by \citet{fontani2011} showed that these 6 sources have integrated line intensities $<$3 K km s$^{-1}$, while the other HMSCs (where DC$_3$N is detected) have 7$-$43 K km s$^{-1}$. 
Another factor that might limit the detection of DC$_3$N in {\it cold} HMSCs is spectral dilution of the line profiles. These cores exhibit narrow linewidths in other deuterated species, e.g. 0.5$-$1.65 km s$^{-1}$ in the N$_2$D$^+$(2-1) transition (\citealt{fontani2011}). These linewidths are similar or slightly larger than the spectral resolution of the observations ($\sim$0.65 km s$^{-1}$), which may produce some dilution of the line intensities. 



The presence of similar levels of deuteration of DC$_3$N in cores with different evolutionary stages does not favor the hypothesis of formation on the surface of dust grains, or at high temperatures during the protostellar phase. 
In these two scenarios, the DC$_3$N emission should be predominantly concentrated around the protostars, and higher values of $D_{\rm frac}$ should be expected in the protostellar phase, due to thermal desorption and hot chemistry, respectively, triggered by the protostellar heating. However, we do not find any preference towards protostellar cores. Moreover, the DC$_3$N map towards the 05358 protocluster (Figure \ref{fig-DC3Nmap}) shows that the emission is extended in the region ($\sim$0.37 pc), with the main deuteration peaks (D1 and D2) shifted with respect to the dust continuum peak where the protostellar activity is on-going. 
Furthermore, hot chemistry in the hot core phase seems unlikely, since it likely too slow to significantly modify the deuteration levels after the desorption of grain mantles (\citealt{charnley1997} and \citealt{osamura2004}).

\begin{figure}
\includegraphics[scale=0.375]{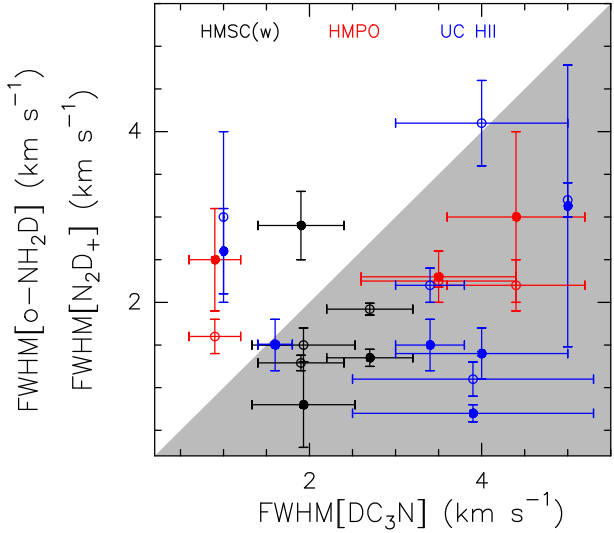}
\centering
\caption{Comparison of the FWHM values of DC$_3$N with those of ortho$-$NH$_2$D(1$_{1,1}-$1$_{0,1}$) (open circles, from \citealt{fontani2015}) and N$_2$D$^+$(2$-$1) (filled circles, from \citealt{fontani2011}). The different colors correspond to different evolutionary groups. The shaded area indicates the parameter space where the FWHM of DC$_3$N is larger than those of the other deuterated species.}
\label{fig-comparison-linewidts}
\end{figure}

Alternatively, the relatively constant level of HC$_3$N deuteration during the different evolutionary phases can be naturally explained if
DC$_3$N mainly arises from material in the outer parts of the clumps, which is common to pre-stellar and protostellar cores. In a region with on-going star formation, 
only the inner gas is heated by the protostar, while the outer envelope remains cold and nearly unaltered from the pre-stellar phase (e.g. \citealt{aikawa2012}). 
The lower values of $D_{\rm frac}$ of HC$_3$N with respect to other species that mainly trace the inner regions of the cores, such as N$_2$H$^+$ or NH$_3$ (Figure \ref{fig-correlations}), points also towards an origin of DC$_3$N in the outer envelope, where the gas is less dense, and hence deuteration is less efficient. 
To further support this hypothesis, we have compared the linewidths of DC$_3$N with those of N$_2$D$^+$ and o$-$NH$_2$D (from \citealt{fontani2011} and \citealt{fontani2015}, respectively). We show in Figure \ref{fig-comparison-linewidts} that the linewidths of DC$_3$N are in most cases larger than those of N$_2$D$^+$ and o$-$NH$_2$D, as expected if the former traces the diffuse outer part of the envelope, and the two latter trace the inner compact regions. We note that the few cases in which N$_2$D$^+$ or o$-$NH$_2$D have larger linewidths than DC$_3$N might be explained by broadening produced by protostellar activity such as molecular outflows. 

In the external part of a star-forming region, the interstellar radiation field (ISRF) is higher, since it is less shielded by the presence of dust. This allows that a higher fraction of carbon is in atomic form, which starts a more rich carbon-chain chemistry. This effect has been detected observationally towards the L1544 pre-stellar core, where many carbon-chain molecules, including HC$_3$N, arise from the external regions of the core, avoiding the inner dust peak (\citealt{spezzano2017}). We have observed a similar behaviour in the 05358 protocluster, where the emission of the $^{13}$C-isotopologues of HC$_3$N, and mainly DC$_3$N, are shifted with respect to the dust continuum (Figure \ref{fig-DC3Nmap}). 

Thus, all the observational evidence presented in this work suggest that DC$_3$N might be a good tracer of deuteration fraction in star-forming regions prior the formation of the denser gas, which is an important parameter in chemical models to constrain, for instance, the ortho-para-H$_2$ ratio (\citealt{pagani2011,kong2015,brunken2014}).
New studies of deuteration of other carbon-chain species (e.g. HC$_5$N, C$_2$H or c-C$_3$H$_2$) in large samples of regions like the one used in this work will serve to better understand how the deuteration of this family of species proceeds during the star formation process.

\begin{figure*}
\includegraphics[scale=0.5]{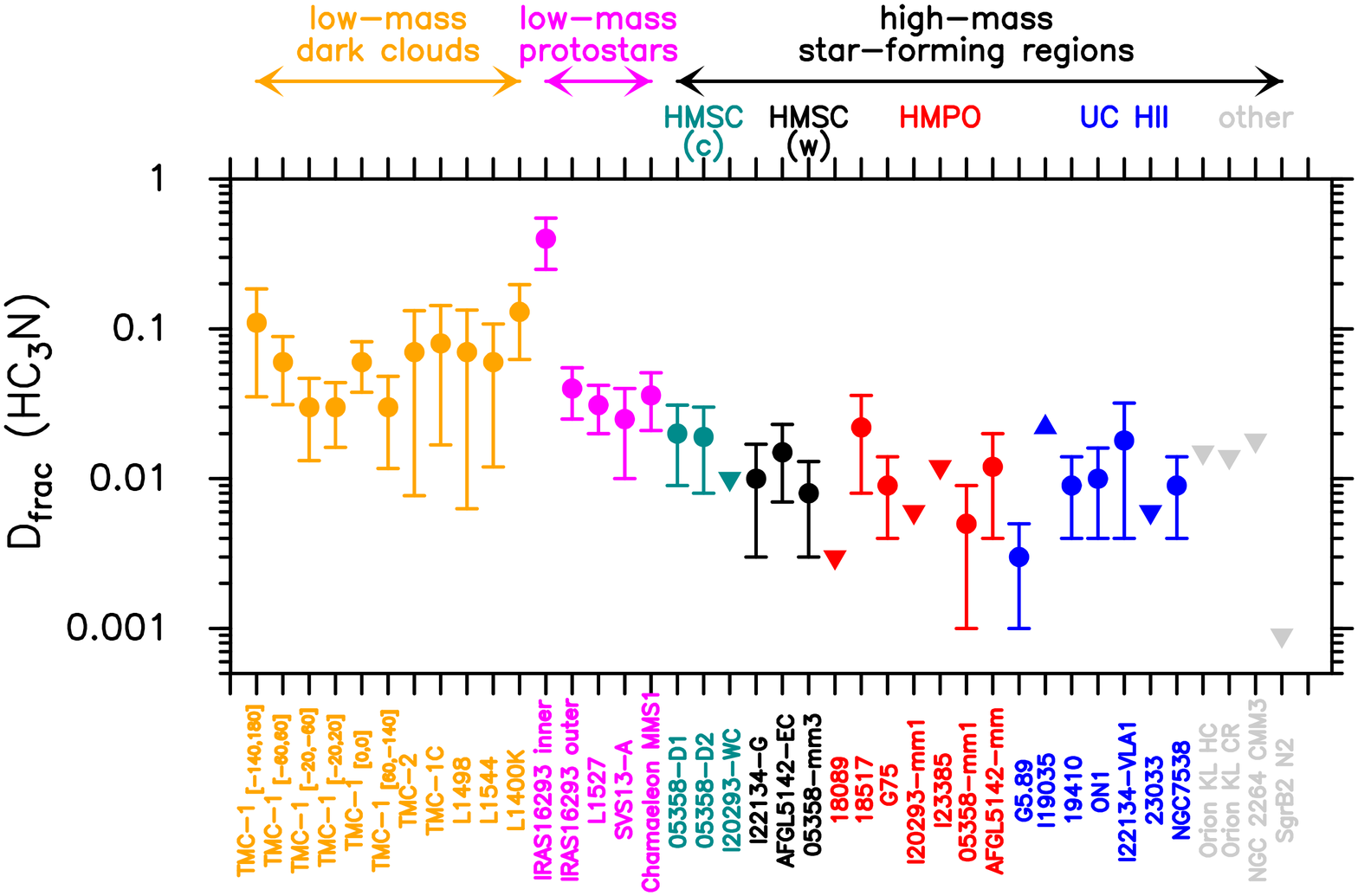}
\centering
\caption{$D_{\rm frac}$ of HC$_3$N in different interstellar sources: low-mass dark clouds from \citet{howe1994} (orange circles); low-mass protostars from \citet{al2017iras,sakai2009,araki2016,bianchi2019,cordiner2012} (magenta circles); high-mass star-forming regions from this work (different colors denote different evolutionary groups, as indicated above the panel; upper limits are denoted by triangles pointing downwards, while the only lower limit is denoted by the triangle pointing upwards); and previous upper limits (gray triangles) from \citet{esplugues2013orion,watanabe2015,belloche2016emoca}.}.
\label{fig-DC3N-all-sources}
\end{figure*}

Finally, in Figure \ref{fig-DC3N-all-sources} we compare the $D_{\rm frac}$ values found in our sample of high-mass cores, divided by evolutionary stage,  with those reported in the literature towards other interstellar sources (low-mass dark clouds, low-mass protostars, and previous upper limits in high-mass star-forming regions; see references in Figure\ref{fig-DC3N-all-sources}). We found values in the interval 0.003$-$0.022 in high-mass cores, which are consistent with the previous upper limits found in Orion KL and SgrB2 N2 hot cores: 0.015 and 0.0009, respectively (\citealt{esplugues2013orion,belloche2016emoca}).
These values of $D_{\rm frac}$ are lower than those found in protostellar low-mass stars (0.025$-$0.5), and pre-stellar low-mass dark clouds (0.03$-$0.13), as shown in Figure \ref{fig-DC3N-all-sources}. This might be due to slightly higher temperatures of high-mass clumps, which would make less efficient the ion-molecule gas-phase reactions responsible of the formation of DC$_3$N.
However, we stress that the comparison between low- and high-mass cores should be done with caution, since the observations are tracing very different linear size scales. As previously mentioned, our observations have likely detected the DC$_3$N arising from the external envelope of the clumps, rather than the inner cores. Therefore, future interferometric observations with high-angular resolution are needed to understand if and how DC$_3$N is formed at smaller core scales ($\sim$0.01 pc).

\section{Conclusions}\label{sec:concl}

We present the first study of DC$_3$N toward a sample of massive cores in different evolutionary stages, from pre-stellar to protostellar phases. We detected the DC$_3$N $J$=11-10 transition towards 15 regions, which include 2 {\it{cold}} High Mass Starless Cores (HMSCs), 3 {\it{warm}} HMSCs, 4 High Mass Protostellar Objects (HMPOs) and 6 Ultra Compact HII regions. We found values of $D_{\rm frac}$ (abundance ratio of DC$_3$N with respect its HC$_3$N) of 0.003$-$0.022, lower than those found in pre-stellar and protostellar low-mass star-forming regions. We do not find any correlation between $D_{\rm frac}$ and the evolutionary stage of the cores, or with the kinetic temperature.
The comparison with other deuterated species previously studied toward the same sample indicates a weak correlation of $D_{\rm frac}$ in those species formed only or predominantly in gas phase (N$_2$H$^+$ and HNC, respectively), and no correlation in species formed only or predominantly on dust grains (CH$_3$OH and NH$_3$, respectively).
We also present the first map of DC$_3$N in a high-mass star-forming region, the protocluster IRAS 05358+3543. The DC$_3$N emission is extended ($\sim$0.37 pc), and it is shifted with respect to the dust continuum peak where the protostellar activity is on-going.
Our observational evidence indicates that DC$_3$N is likely formed by gas-phase ion-molecule reactions in the outer and less dense part of star-forming clumps, where the interstellar radiation field keeps most of the carbon in atomic form, enhancing the formation of carbon-chain molecules such as HC$_3$N. Thus, DC$_3$N might be a good tracer of the deuteration level in star-forming regions prior to the formation of denser gas.


\section{Acknowledgements}

\noindent
We acknowledge the anonymous reviewer for her/his careful reading of the manuscript and her/his useful comments.
V.M.R. has received funding from the European Union's Horizon 2020 research and innovation programme under the Marie Sk\l{}odowska-Curie grant agreement No 664931. LC acknowledges support from the Italian Ministero dell’Istruzione, Universit\`a e Ricerca through the grant Progetti Premiali 2012 - iALMA (CUP C52I13000140001). This work is based on observations carried out under projects number 129-12 and 040-19 with the IRAM 30m telescope. IRAM is supported by INSU/CNRS (France), MPG (Germany) and IGN (Spain). This publication has received funding from the European Union’s Horizon 2020 research and innovation programme under grant agreement No 730562 [RadioNet]. L.C. and F.F. thank the IRAM staff for the precious help provided during the observations.
We warmly thank Xing "Walker" Lu and Qizhou Zhang for sharing with us their VLA map of the 05358+3543 high-mass protocluster.




\bibliographystyle{mnras}
\bibliography{bibfile.bib} 






\bsp	
\label{lastpage}
\end{document}